# CRM 2.0 within E-Health Systems:

## Towards Achieving Health Literacy & Customer Satisfaction

Muhammad Anshari
PhD Cand. at Universiti Brunei Darussalam, &
Department of Informatics, UIN Yogyakarta
anshari@yahoo.com

Mohammad Nabil Almunawar
FBEPS-UBD, Jl.Tungku Link Gadong,
BE 1410 Brunei Darussalam
nabil.almunawar@ubd.edu.bn

Patrick Kim Cheng Low
FBEPS-UBD, Jl.Tungku Link Gadong,
BE 1410 Brunei Darussalam
patrick.low@ubd.edu.bn

*Abstract*—Customer Relationship Management (CRM) within healthcare organization can be viewed as a strategy to attract new customers and retaining them throughout their entire lifetime of relationships. At the same time, the advancement of Web technology known as Web 2.0 plays a significant part in the CRM transition which drives social change that impacts all institutions including business and healthcare organizations. This new paradigm has been named as Social CRM or CRM 2.0 because it is based on Web 2.0. We conducted survey to examine the features of CRM 2.0 in healthcare scenario to the customer in Brunei Darussalam. We draw the conclusion that the CRM 2.0 in healthcare technologies has brought a possibility to extend the services of e-health by enabling patients, patient's families, and community at large to participate more actively in the process of health education; it helps improve health literacy through empowerment, social networking process, and online health educator. This paper is based on our works presented at ICID 2011.

*Keywords-CRM 2.0; Brunei Darussalam; Health Literacy; Customer Satisfaction; Healthcare Organization*

## I. INTRODUCTION

Since the study of e-health is considered as multidisciplinary knowledge, many topics on e-health have been partially overviewed from different domain of expertise. Those domains of expertise are technology, people (culture), and process (strategies, effects, and outcome). Therefore, a challenging task is to assemble all the interrelated disciplines together into a framework that can be used to improve health literacy and customer satisfaction at the same time.

The term e-health is an emerging field as the intersection of healthcare, Information Communication Technology, and business process, referring to health services and information delivered or enhanced through the Internet and related technologies. In a broader sense, the term characterizes not only a technical development, but also a state-of-mind, a way of thinking, an attitude, and a commitment for networked, global thinking, to improve health care locally, regionally, and worldwide by using information and communication technology [1].

Marconi (2008) defines e-health as an application of Internet and other related technologies in the healthcare industry to improve access, efficiency, effectiveness, and quality of clinical and business processes utilized by healthcare organizations, practitioners, patients, and consumers in an effort to improve the health status of patients [2]. E-health can provide patient with opportunities to interact with their healthcare systems online and can provide new forms of patient-physician interaction [3].

Samples of e-health application are Patient-centered Health Record (PcHR) that was created by research group at the University of Washington. Using the PcHR, the patient's own record of their health information was integrated into clinical scenarios that involved exchanging health records with traditional clinical information systems [4]. Lee Ting, et.al. (2006) developed patient-oriented diabetic education management (POEM) system. The system can provide the incentives for patients to continuously and persistently log in and learn the required knowledge and skills, improving their clinical outcomes [5].

Meanwhile, one of the most interesting aspects in e-health is how to manage the relationship between healthcare providers and patients. As in nature, fostering relationship leads to maintain loyal customer, greater mutual understanding, trust, patient satisfaction, and patient involvement in decision making [6]. Furthermore, effective communication is often associated with improved physical health, more effective chronic disease management, and better health related quality of life [7]. On the other hand, failure in managing the relationship will affect to the patient dissatisfaction, distrust towards systems, patient feels alienated in the hospital, and jeopardize business survivability in the future.

Therefore, adopting CRM within e-health can be viewed as strategy to attract new customers coming to an organization, retaining them throughout the entire lifetime of a relationship, and extending other services or products to the existing customers. In the healthcare environment, healthcare providers are challenged to acquire potential customers for the healthcare services, retaining them to use the services, and extending various services in the future. To take the challenges, healthcare provider must consider establishing close of relationship with their patients offer convenience of services, and provide transparency in services through information sharing.

The main goal of this paper is to introduce a promising future research direction which will shape the future of health informatics. In this paper we will discuss how the new approach will help the healthcare increasing their customer support, avoiding conflict, and promoting better health literacy to patient, and improve customer satisfaction. The case of Brunei Darussalam (Brunei) will be presented in term of e-health's perception.

The structure of the paper is as follows. In section 2, we discuss CRM, e-health technologies, and customer service in healthcare. In section 3, we focus on e-health perception in Brunei. We discuss intersection of CRM 2.0 within e-health system in section 4, and finally section 5 is the conclusion.

## II. BACKGROUND OF STUDY

Today, every healthcare organization depends on ICT in every level of activities. The healthcare organization relies on process application and information streamline to create value for every facet of its delivery. The use of ICT in healthcare organizations has grown in the same pattern as compare to any other industry landscape. The use of web technology, database management systems, and internetwork infrastructure are part of ICT initiative that will affect of healthcare practice and administration. We propose our conceptual idea CRM 2.0 or Social CRM as extension of CRM with the Web 2.0 capabilities in healthcare organizations. The idea operates in the area of healthcare provider–patient and patient-patient relationships inclusive with social networks interaction, and how they possibly shared information to achieve health outcomes. It offers a starting point for identifying possible theoretical mechanisms that might account for ways in which CRM 2.0 provides an alternative for building active relationship between healthcare provider, patients, and community at large. This section is to briefly summarize the current health information systems and to identify few emerging trends and research in e-health

### A. CRM

In practice, many see CRM as merely a technology for improving customer service which may lead to a failure when implementing it. CRM initiatives must be seen as a strategy for significant improvement in services by solidifying satisfaction, loyalty and advocacy through information and communication technology. As such, matters pertaining related to people such as customer behaviours, culture transformation, personal agendas, and new interactions between individuals and group must be incorporated in CRM initiatives. Therefore, organization needs to understand that behaviours and expectations of customers (patients in healthcare organizations) which continue to change overtime. Consequently, CRM must address the dynamic nature of patients' needs and hence adjustments strategies embedded in CRM are required.

Greenberg (2009) defined CRM as a philosophy and a business strategy supported by a system and a technology designed to improve human interactions in a business environment. It is an operational, transactional approach to customer management focusing around the customer facing departments, sales, marketing and customer service [8].

Early CRM initiatives was the process of modification, culture change, technology and automation through use of data to support the management of customers so it can meet a business value of corporate objectives such as increases in revenue, higher margins, increase in selling time, campaign effectiveness, and reduction in call queuing time. While nowadays, CRM is designed to engage the customer in a collaborative conversation in order to provide mutually beneficial value in a trusted and transparent business environment.

Currently, a new paradigm has appeared in CRM systems as a result of the development of IT and Web service. This new paradigm has been named as Social CRM or CRM 2.0 because it is based on Web 2.0 [8].

CRM has evolved people, processes, and contents which transform from customer management to customer engagement [8]. That is more focused on the conversation that is going on between healthcare-patient and patients-others. Web 2.0, which play a significant part in the CRM transformation drives social change that impacts all institution including business and healthcare organizations. It is a revolution on how people communicate. It facilitates peer-to-peer collaboration and easy access to real time communication. Because much of the communication transition is organized around web based technologies, it is called Web 2.0 [8]. Patients participate in these social network can share information about their diagnoses, medications, healthcare experiences, and other information. It is often in form of unstructured communication which can provide new insights for people involved in the management of health status and chronic care conditions.

The term of Social CRM and CRM 2.0 is used interchangeably. Both share new special capabilities of social media and social networks that provide powerful new approaches to surpass traditional CRM. Greenberg (2009) defined Social CRM as a philosophy and a business strategy, supported by a technology platform, business rules, processes, and social characteristics, designed to engage the customer in a collaborative conversation in order to provide mutually beneficial value in a trusted and transparent business environment. It's the company's response to the customer's ownership of the conversation [8]. Fabio Cipriani described the fundamental changes that Social CRM is introducing to the current, traditional CRM in term of landscape. Figure 1 is reflection of the evolving CRM 2.0 which is different from CRM 1.0. It is a revolution in how people communicate, customers establish conversation not only with the service provider but it is also with others [9].

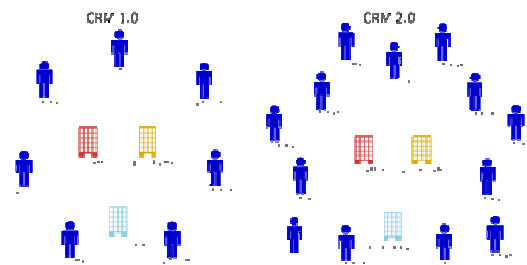

Fig. 1; Evolution of CRM landscape (Fabio Capriani)

### B. E-Health

The vision of a paperless hospital is delineated as the embodiment of the future health information systems with the hope is that brings an improvement with promise of to be more reliable effective and efficient. The current status of HIS varies among countries. There are 193 countries that are a member of World Health Organization (WHO) in 2009; 114 of them participated in the global survey on e-health [10]. Most developed countries have fully utilized HIS in their systems as they have the resources, expertise, and capital to implement them while in the developing countries, HIS have not been fully utilized yet.

One such trend in HIS is the slowly adoption of e-health systems toward the use of EMR. The systems move patient information from paper to electronic file formats so they can be easily and effectively managed. However, an interesting fact to be noted that the tendency of people to know more and actively participate in the health promotion, prevention, and care together with the rights that will become a standard legislature guide the development of information systems that support these tendencies. Thus, the trend is towards more patients centred or involvement of patients in receiving information, in decision making and in responsibility for their own health. The main feature of this trend is the shifting from healthcare-institution cantered care to the patient-centred care emphasizing on continuity of care from prevention to rehabilitation. This objective can be achieved through shared care process which builds on health telematics networks and services, linking health providers, laboratories, pharmacies, insurance company and social centres offering to individuals a 'virtual healthcare centre with a single point of entry. Furthermore, this vision implies provision of health services to homes with innovative services such as personal health monitoring and support systems and user-friendly information systems for supporting health education and awareness [11].

Several examples of countries that have implemented HIS are Canada, Singapore and Australia to name a few. Canada has established e-health Ontario on March 2009 with three targeted strategies to improve; diabetes management, medication management and wait times. One of the examples of the service offered is ePrescribing under the medication management. It authorizes and transmits prescriptions from physicians and other prescribers to pharmacists and other dispensers [12]. It prevents medication error due to illegible prescribing and reduces fraud prescriptions. "Participating prescribers and pharmacies at both sites will continue electronically prescribing until a provincial Medication Management System is in place" [12]. While, Australia will release personally controlled electronic health records (PCEHR) on July 2012. The system will enable better access to important health information held in dispersed records across the country. For the first time all Australians who choose to participate will be able to see their important health information when and where they need it, and they will be able to share this information with trusted healthcare providers [13].

An emerging trend in HIS is the use of web technologies. The benefits of web services include ease of integration and ease of accessibility. There are a number of health information systems demonstrating applications of computer networks that tap into the vast array of health information available on the web. These systems are designed for patients with a health crisis or medical concern and for primary care providers. Patient-centred health information systems include models that use a variety of strategies to filter health information from the web to provide focused and tailored health information for concerns such as HIV/AIDS, asthma, smoking cessation, living with alcohol abuse, or stress [14,15]. Physician-centred health information support systems use a variety of strategies to provide health information filtered from the WWW for physicians and patients/families on topics such as child health, safety, and the management of insulin-dependent diabetes [16, 17]. A HealthGrid allows the gathering and sharing of many medical, health and clinical records/databanks maintained by disparate hospitals, health organizations, and drug companies. In other words, HealthGrid is an environment in which data of medical interest can be stored and made easily available to different actors in the healthcare system, physicians, allied professions, healthcare centres, administrators and, of course, patients and citizens in general [18]. Also, the driving forces for the individual and commercial adoption of the VoIP are the significant cost savings, portability, and functionality that can be realized by switching some or all of their voice services to VoIP. Chen et al showed the integration of mobile health information system with VoIP technology in a wireless hospital [19].

From technological perspective, there are various emerging tools and technologies in creating and managing HIS. Semantic Web is an extension of the World Wide Web, offers a united approach to knowledge management and information processing by using standards to represent machine-interpretable information. Semantic Web technology helps computers and people to work better together by giving the contents well-defined meanings. The semantic Web has also drawn attention in the medical research communities [20]. Semantic web services can support a service description language that can be used to enable an intelligent agent to behave more like a human user in locating suitable Web services. While, Web services are software components or applications, which interact using open XML and Internet technologies. These technologies are used for expressing application logic and information, and for transporting information as messages [21]. They have significantly increased interest in Service oriented architectures (SOAs) [22].

Recent development in ubiquitous computing is a paradigm shift since technology becomes virtually invisible in our lives. The ubiquitous computing environment will make possible new forms of organizing, communicating, working and living. However, ubiquitous computing systems create new risks to security and privacy. To organize the u-healthcare infrastructure, it is necessary to establish a context-aware framework appropriate for the wearable computer or small-sized portable personal computer in ubiquitous environment [23]. The mobile health (m-health) a form of ubiquitous computing can be defined as mobile communications network technologies for healthcare [24]. This concept represents the evolution of "traditional" e-health systems from desktop platforms and wired connections to the use of more compact devices and wireless connections in e-health systems.

The next emerging trend in HIS is to use Web 2.0 technologies, which use the Internet as a working platform. Recently, the Web 2.0 tools such as Facebook, Twitter, Myspace, Friendster, LinkedIn, etc. have grown rapidly facilitating peer-to-peer collaboration, ease of participation, and ease of networking. The use of Web 2.0 in HIS system is equivalent to bringing patient expectation aligned with fashion of ICT in actual healthcare services. It offers new outlook either from patient or healthcare organization, and how they structure inter-relation between three distinct domains of objects; customer's expectation, advancement of ICT, and healthcare services. Each domain has unique features and characteristics which failing to respond appropriately may affect to business survivability and customer dissatisfaction. The main advantages of Web 2.0 are the linkage among people, ideas, processes, systems, contents and other organizational activities [25]. Therefore, Web 2.0 could affect healthcare business process like relationship between patients and healthcare providers as it is about engaging relationships, sharing experience and information, and collaboration. However, the effects of Web 2.0, particularly in addressing the issues of healthcare services are still not much discussed.

Web 2.0 technologies has brought a possibility to extend the service of HIS by enabling patients, patient's families, and community at large to participate more actively in the process of health promotion and education through social networking process. We proposes integrated HIS which includes holistic approach of personal habit, physical activities, spiritual and emotional activities, and social support as well as social networks to be part of the systems. It is significant to improve customer satisfaction and health literacy in healthcare service in order to accommodate components and features of social networking capabilities, empowering patients, and availability of online health educator [26].

*C. Customer Service & Customer Satisfaction in Healthcare*

As a business, healthcare organization stands in need of the same standards of customer service as other industries or business organizations. The fact that customer service expectations in healthcare organization are high poses a serious challenge for healthcare providers as they have to make exceptional impression on every customer. In the competitive commercial healthcare market, poor service leads customers to switch healthcare providers because poor service indicates inefficiency, higher cost and lower quality of care. One of four Americans has switched or considered switching doctors (26 percent) or hospitals or clinics (23 percent) because of negative experiences. More than half (52 percent) say they choose hospitals and clinics based on whether they believe employees understand their needs [27]. The report also recommends health care providers become "empathy engines," that is transforming their organizations to allow frontline employees to focus on patient problems and innovate. This applies to hospitals, clinics, payers, vendors and pharmacy chains as well.

Nowadays, more patients have more choices in where they seek care and how they interact with their healthcare providers. A great customer service can lead to major improvements in the health care system. Customer service is not an "extra"— it is an essential requirement for providing high quality healthcare and for staying in highly competitive business [28]. Patients are making clear choices about where they receive care based on service experiences and it is crucial for organizations to create an institutional ability to sense and respond empathetically [27].

For instance, it will give more options to patient for consultation either physically meet the doctor or to make it online as shown in figure 2. The figure explains the state of the personal/patient health cycle. It consists of five condition, these are personal daily life which is daily habitual that may affect personal health. Checkups are the condition where patient needs to visit medical staffs for assessment or consultation. Outpatient treatment is condition where patient in state of health monitoring while considered not necessary to stay at hospital. Inpatient treatment is condition which requires patient has to stay at hospital for intensive treatment. And finally, social life is condition where patient will likely to discuss and share their health condition with friends or relatives. There are condition that e-health can contribute in each and every process of personal health cycle.

For instance, in the state of personal daily life and social life (see figure 2) yet individual may carry on to have interaction in the process of healthcare services between healthcare provider-patient, and patient-patient communication. The interactions enable them to improve their health literacy.

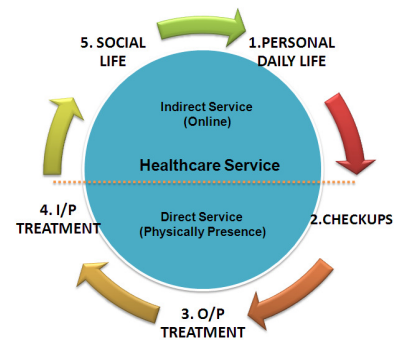

Figure 2. E-health and Healthcare Cycle

Healthcare organization strategies should transform customer strategies and systems to customer engagement. The one is more focused on the conversation that is going on between organization and customer and the collaborative models that cutting edge companies are carrying out for customer engagement. Proactive strategies will improve customer services. And great customer support will increase loyalty, revenue, brand recognition, and business opportunity.

*D. Health Literacy*

Health literacy is being increasingly recognized as important to health and health outcomes [29]. Health literacy is not simply the ability to read health information. It is a learning process which will improve regularly when the patients get used to it and requires a process of reading, listening, analytical, and decision-making skills, and act on medical instructions. It includes the ability to understand instructions on prescription drug bottles, appointment slips, medical education brochures, doctor's directions and consent forms, and the ability to negotiate complex health care systems.

The definition of health literacy first proposed by Ratzan and Parker (2000) has generated many discussions [30]. It is broadening the idea that communication is more than the information that is put out—it is also about what individuals take in. Health literacy reflects the dual nature of communication: what information is being disseminated and how people understand the information given to them [31]. Looking at health literacy in this way enables a focus on making systematic improvements rather than blaming those with low health literacy skills.

Some definitions of literacy vary from a simplistic understanding of literacy as the capacity to read, write, and have basic numeric skills to one that considers for complexity, culture, individual empowerment, and community development [32]. A working definition of literacy that attempts to integrate a broad range of factors has been proposed by the Center for Literacy of Quebec; Literacy involves a complex set of abilities to understand and use the dominant symbol system of a culture for personal and community development [33]. The need and demand for these abilities vary in different societies. In a technological society, the concept is expending to include the media and electronic text in addition to alphabets and numbers. Individuals must be given life-long learning opportunities to move along a continuum that includes reading, writing, critical understanding and decision making abilities they need in their communities.

Poor health literacy is common among racial and ethnic minorities, elderly persons, and patients with chronic conditions, particularly in public-sector settings. Studies suggest that while individuals with limited health literacy come from many group of people, the problem of limited health literacy is often greater among older adults, people with limited education, and those with limited English proficiency [34, 35, 36]. Limited health literacy program come from many group of people, but the advancement of web technology gives impact to all sectors including healthcare organization. The study is to examine among people in Brunei the expectation of e-health initiative in order to improve their health literacy.

### III. E-HEALTH PERCEPTION

To give better idea of the relationship between CRM 2.0 within e-health, we conducted survey to public at major hospitals, clinics, or home care centres around Brunei starting from February to March, 2011. E-health's expectation in relation with health literacy was measured by descriptive analysis. It is purposive sampling methods in which they were intentionally selected from patients, patient's family, or medical staffs from hospitals, clinics, and homecare centre across the country. Majority of the respondents are local Bruneians, who frequently visit health practitioners. The respondents range from twenty years old (or younger) to above 51 years old. Therefore, they represent a fair share of the general public. There were 366 respondents participating for the survey and conducted from February to March 2011. Because no research has ever been published on e-health and health information accessibility issues in Brunei, this paper is prepared to fill that gap. Many e-health topics are covered, including existing health information systems, health information accessibility, and current issues like social support and social networking in healthcare organization. Data gathered from the survey will be used to formulate recommendation for future direction of e-health systems.

In order to know a coefficient of reliability (consistency) of questionnaire, the authors use Cronbach's alpha to measure of internal consistency, that is, how to closely related set of items are as a group.

$$\alpha = \frac{N.\bar{c}}{\bar{v} + (N-1).\bar{c}} \quad (1)$$

Here $N$ is equal to the number of items, c-bar is the average inter-item covariance among the items and v-bar equals the average variance (1). If Cronbach's alpha is greater than or equal to 0.6 then a variable is reliable.

**TABLE 1**
COEFFICIENT OF RELIABILITY

|       |          | N   | %     |
|-------|----------|-----|-------|
| Cases | Valid    | 363 | 99.2  |
|       | Excluded | 3   | .8    |
|       | Total    | 366 | 100.0 |

From the table 1 shows that Cronbach's Alpha is found 0.80 from the 363 number of items. It is indicated that they have relatively high internal consistency since it is 0.80 > 0.60, and the survey result is reliable.

Patients' functional health literacy strongly correlated with supportive and availability of health system. For example, e-health services will eventually make an organization more efficient and effective in managing its resources hence leading to greater productivity. Electronic records make it easier to schedule appointments for patients, keep track of follow-ups, and ensure patients' general practitioners are informed of the results of their referrals. Indeed, 82% of respondents agreed to use the service if it is available. In addition, majority of respondents (79%) agreed to view and deal with their own health record online. Viewing medical record online is also features that will encourage patient to be proactive with health promotion, boost healthcare awareness, and self managed healthcare.

TABLE 1 e-health with CRM 2.0 and Point Intervention

| Point Intervention | e-health's features enable to |
|---|---|
|  |  |

---

Based on our works presented on ICID2011 [1]

| | |
|---|---|
| **Culture & Society** | 1. Discuss about health services in social networks<br>2. Discuss about health status with friends in social networks<br>3. Online support/sharing with other patient who have similar condition<br>4. Use social networks/online support group provided by healthcare provider |
| **Health System** | 1. Make appointment online<br>2. View medical records online<br>3. Request refills prescriptions online<br>4. View medical payment online<br>5. Control with whom I share my online medical records |
| **Education System** | 1. Level of education<br>2. Age composition<br>3. Internet access<br>4. Interact with online health promotion program<br>5. Use ICT to consult with health educator |

Table 1 shows overall levels of e-health features in point interventions with survey result were demonstrated. It is statistically significant in scores on the general of point intervention were noted between culture and society, health system, and education system. The intervention of health literacy program should highly consider to the adoption of social networks within healthcare organization, availability of online support to patients, and media sharing among patients who have similar condition and symptoms. In terms of health systems, healthcare organization must respond demands of patients towards e-health services such as make online appointment, give online access to their medical records, view medical payment online (if necessary), and give them empowerment with whom they want to share their medical records. Last but not least, education system that affects health literacy, health literacy program should accommodate all level of education and ages appropriately based on their knowledge level and interest, availability of Internet access to all targeted audience, and ability to use ICT in dealing with online health education.

## IV. DISCUSSION

CRM can be viewed as strategy to attract new customers coming to an organization, retaining them throughout the entire lifetime of a relationship, and extending other services or products to the existing customers. In the healthcare environment, healthcare organizations are challenged to acquire potential customers for the healthcare services, retaining them to use the services, and extending various services in the future. To take the challenges, healthcare organization must consider establishing close of relationship with their patients offer convenience of services, and provide transparency in services through information sharing.

Therefore, the healthcare organization should perform re-engineering process to adapt their CRM strategy and tool in order to acquire potential customer coming for the service [37]. Though e-health is not going to replace whole existing healthcare services, it is extendable systems which features and services can improve quality of service, provide effectiveness in process, and convenience for patient. It also will form healthcare service more comprehensive and reliable in serving patients.

From the survey, we found that expectations in healthcare services are high, which create challenge for healthcare organization. In response to the expectation, it is significant to add features as front end to tie up the interface of e-health. From the survey, it is important to improve customer satisfaction and health literacy in healthcare service to accommodate components and features of social networking capabilities, empowering patients, and availability of online health educator (see figure 3).

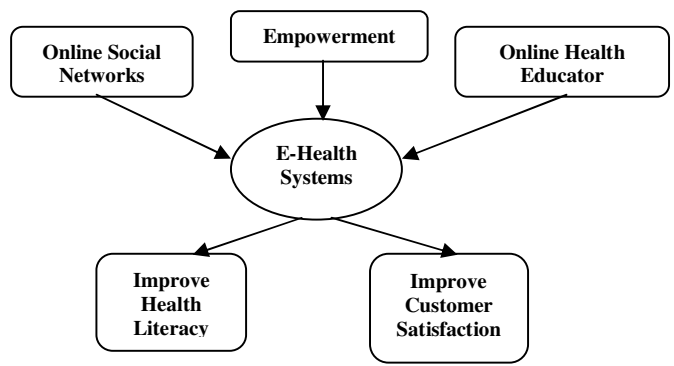

Figure 3. E-health with features based on Survey

The survey confirmed that patients preferred to have more ability and control of health information and application that concern with their health status. Making appointment online, viewing their medical records, and consulting online with medical staffs are few examples of empowerment to patient. The authorization to access health information will benefit both patient and healthcare organization. From patient perspective, their health literacy is expected to improve by the time they have better knowledge of their own health status, and for the healthcare organization is expected to be long lasting relationship since they always in need to access the service. Additionally, empowerment is also believed giving them flexible time when and where they want to look upon their health related activities.

The majority of respondents agreed that social networks and social support online part of the e-health systems. It should be operated, managed, and maintained within healthcare's infrastructure. This is more targeted to internal patients/families within the healthcare to have conversation between patients/family within the same interest or health problem/ illness. For example, patient with diabetic would motivate to share his/her experiences, learning, and knowledge with other diabetic patients. Since patient/family who generates the contents of the Web, it can promote useful learning centre for others, not only promoting health among each others, but also it could be the best place supporting group and sharing their experiences related to all issues such as; how the healthcare does a treatment, how much it will cost them, what insurance accepted by healthcare, how is the food and nutrition provided, etc. Furthermore, it is also suggested that medical staff or doctors should be actively involved in an online forum discussion to provide professional advice. For healthcare management, conversations generate between patients in social networks site can be starting place to construct business strategy for the organization. Social networking site should be moderated by competent healthcare staff within the managerial level to capture the conversation and listen what patients say about the service. The role of this task could campaign of healthy life for the society which is not intended for commercial benefit for short term but it is beneficial for the community.

In light of this, from the survey, confirmed that the availability of online health educator is important. Presence of online health educator is the vital point in e-health instead of ICT as tool. They are expected to have skills to interpret medical data, able to guide patient go through technical systems, and know how to respond online queries properly.

## V. CONCLUSION

In a competitive commercial healthcare environment, negative experience and poor service lead customers to switch healthcare providers because poor service indicates inefficiency, higher cost and lower quality of care. The high expectation of customer service provided by healthcare organizations in the information age poses a serious challenge for healthcare providers as they have to make an exceptional impression on every customer. The adoption of CRM 2.0 features as front end to tie up the interface of e-health is believed to boost effectiveness and efficiency customer service in healthcare organizations. From the survey, we found that expectations in healthcare services are high, which create challenge for healthcare organization. Furthermore, it is significant to improve customer satisfaction and health literacy in healthcare service at least e-health systems should accommodates components and features of social networking capabilities, empowering patients, and availability of online health educator.

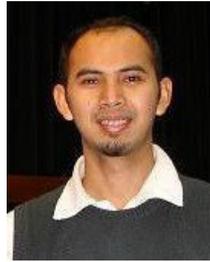

**AUTHORS PROFILE**

**Muhammad Anshari,** is a business information system's practitioner and researcher. He received his BMIS (Hons) from International Islamic University Malaysia, his Master of IT from James Cook University Australia, and currently, he is pursuing his PhD program at Universiti Brunei Darussalam. His professional experience started when he spent two years as an IT Business Analyst at PT. Astra International Tbk. (National Automotive Company), joined research collaboration on *Halal* Information Systems at Chulalongkorn University Bangkok - Thailand, and researcher at College of Computer and Information Science King Saud University, Riyadh-Saudi Arabia. His main research interest is on CRM, ERP, SCM, and healthcare management issues in relation with people, process and ICT. His email address is anshari@yahoo.com.

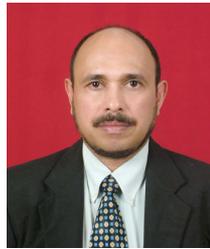

**Mohammad Nabil Almunawar** Ir. IPB Indonesia, MSc UWO Canada, Ph.D. UNSW, Australia is a senior lecturer and the Deputy Dean (Graduate Studies and Research) at Faculty of Business, Economics and Policy Studies, Universiti of Brunei Darussalam (UBD), Brunei Darussalam. Dr Almunawar has published many papers in refereed journals as well as international conferences. He has many years teaching experiences and consultancies in the area computer and information systems. His overall research interest is application of IT in Management and e-commerce/e-business/e-government. He is also interested in object-oriented technology, multimedia information retrieval, health information systems and information security.

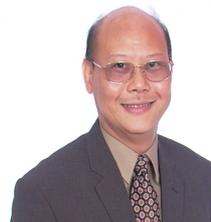

**Prof. Dr. Patrick Kim Cheng Low, Ph.D.** (South Australia), Chartered Marketer, Certified MBTI Administrator, & Certified Behavioral Consultant (IML, USA), brings with him more than 21 years of combined experience from sectors as diverse as the electronics, civil service, academia, banking, human resource development and consulting. The once Visiting Professor, Graduate School of Business, the University of Malaya (Jan to Feb 2007), Prof. Dr. Low was the Deputy Dean, Postgraduate Studies & Research, teaching in Universiti Brunei Darussalam (2009). He teaches the graduate students/ MBA in Organizational Behavior, Managing Negotiations, Leadership and Change Management, and the undergraduates in Leadership Basics, Challenging Leadership, Business and Society, Issues in Organizational Leadership, Organization Analysis & Design; and Organization Development & Change. The former Associate Dean, Director of Career Services and Chair of the Management and Marketing Department of a University in Kazakhstan (2004 to 2006) focuses on human resource management and behavioral skills training covering areas like negotiation/ influencing, leadership and behavioral modification. He can be contacted at patrick_low2003@yahoo.com.